\documentclass[fleqn,usenatbib]{mnras}

\hypersetup{citecolor=blue,urlcolor=magenta}

\usepackage[T1]{fontenc}
\usepackage{booktabs}
\usepackage{graphicx}
\usepackage [caption=false]{subfig}

\DeclareRobustCommand{\VAN}[3]{#2}
\let\VANthebibliography\thebibliography
\def\thebibliography{\DeclareRobustCommand{\VAN}[3]{##3}\VANthebibliography}
\newcommand\ec{$E_{\rm cut} $ }

\usepackage{graphicx}	
\usepackage{amsmath}	
\usepackage{amssymb}	

\usepackage{newtxtext,newtxmath}
\usepackage{xcolor}

\title[X-ray high-energy cutoff in Mrk 348]{The X-ray high-energy cutoff in Compact Symmetric Object Mrk 348}

\author[M. Liao et al.]{
Mai Liao$^{1,2,3}$\thanks{mai.laio@mail.udp.cl and liaomai@ustc.edu.cn},
Junxian Wang$^{4,5}$\thanks{jxw@ustc.edu.cn},
Jialai Kang$^{4,5}$\thanks{ericofk@mail.ustc.edu.cn},
Xiaofeng Li$^{6,7,8}$,
Minhua Zhou$^{9}$
\\
$^{1}$National Astronomical Observatories, Chinese Academy of Sciences, 20A Datun Road, Chaoyang District, Beijing 100101, China\\
$^{2}$Chinese Academy of Sciences South America Center for Astronomy, National Astronomical Observatories, CAS, Beijing, 100101, China\\
$^{3}$Instituto de Estudios Astrofísicos Facultad de Ingeniería y Ciencias Universidad Diego Portales Av. Ejército 441, Santiago, Chile\\
$^{4}$CAS Key Laboratory for Research in Galaxies and Cosmology, Department of Astronomy, University of Science and Technology of China, Hefei, Anhui 230026, China\\
$^{5}$School of Astronomy and Space Science, University of Science and Technology of China, Hefei 230026, China\\
$^{6}$Department of Astronomy, Guangzhou University, Guangzhou 510006, PR China\\
$^{7}$Great Bay Center, National Astronomical Data Center, Guangzhou, Guangdong 510006, China\\
$^{8}$Astronomy Science and Technology Research Laboratory of Department of Education of Guangdong Province, Guangzhou 510006, China\\
$^{9}$School of Physical Science and Intelligent Education, Shangrao Normal University, 401 Zhimin Road, Shangrao 334001, China
}

\pubyear{2023}

\begin{document}
\label{firstpage}
\pagerange{\pageref{firstpage}--\pageref{lastpage}}
\maketitle

\begin{abstract}
Compact radio AGN are thought to be young radio active galactic nuclei (AGN) at the early stage of AGN evolution, thus are ideal laboratory to study the high-energy emission throughout the evolution of radio AGN.
In this work, we report for the first time the detection of the high-energy cutoff ($E_{\rm cut}$), a direct indicator of thermal coronal radiation, of X-ray emission in Mrk 348 ($z$ = 0.015), a young radio galaxy classified as compact symmetric object.
With a 100 ks NuSTAR exposure, we find that the high-energy cutoff ($E_{\rm cut}$ ) is firmly detected ($218^{+124}_{-62}$ keV). Fitting with various Comptonization models indicates the presence of a hot corona with temperature $kT_{\rm e}$ = 
35 -- 40 keV. These strongly support the corona origin for its hard X-ray emission.
The comparison in the $E_{\rm cut}$ -- spectra index $\Gamma$ plot of Mrk 348 with normal large-scale radio galaxies (mostly FR II) yields no difference between them. This suggests the corona properties in radio sources may not evolve over time (i.e., from the infant stage to mature stage), which is to-be-confirmed with future sample studies of young radio AGN.

\end{abstract}




\section{Introduction}
Gigahertz-peaked spectrum (GPS), compact steep spectrum (CSS), and high-frequency peaked (HFP) radio galaxies and quasars are a subclass of radio-loud AGN characterized by an intrinsically compact radio morphology and a peaked radio spectrum. Their peak frequencies in radio spectra are around 1 GHz, 100 MHz, and above 5 GHz for GPS, CSS, and HFP, respectively. The linear size (LS) in both GPS and HFP sources are below 1 kpc, while in CSS sources well confined within 20 kpc.
Such radio galaxies typically have small classical double radio structure, namely, Compact Symmetric Objects CSOs (LS $\le 1$ kpc) and Medium Symmetric Objects MSOs ($\le$ 20 kpc), and type II optical spectrum. Meanwhile these quasars identified with Type I spectra tend to show evidence of beaming and have a core-jet radio morphology \citep[see the review of][]{2021A&ARv..29....3O}.


At present, it is generally believed that these compact radio AGN are at the early stage of AGN evolution, inferred by revealing scaled-down morphologies reminiscent of large-scale radio lobes of powerful radio galaxies FR II/ FR I \citep{1974MNRAS.167P..31F} from the very long baseline interferometry (VLBI) observations. They are thus hypothesised to be the precursors to the classical large-scale radio galaxies \citep{2003PASA...20...69P,2012ApJ...760...77A}. 
The "youth scenario" is further strongly supported by the measurements of dynamical age for a few dozens of CSOs \citep[about $10^{2}$ $-$ $10^{4}$ yr, e.g.,][]{2012ApJS..198....5A}, and by the spectral age ($10^{3}$ $-$ $10^{5}$ yr) from the high-frequency spectral break modeling \citep[e.g.,][]{2009AN....330..193G}. Hence, compact radio AGN (hereafter young radio AGN) are of great interest in their own,
and 
are conducive 
to studying many core issues of AGN, e.g., uncovering the radio/accretion activity at the initial stage of AGN, understanding the evolution of AGN, and probing high energy radiative processes dominating the X-ray/$\gamma$-ray output throughout the evolution of AGN.   


Although majority of the works up to now on young radio AGN focus on their radio properties, studies on their X-ray emission also have been significantly increasing in recent years, mainly with the observations of the Chandra and XMM-Newton \citep[e.g.,][]{2006A&A...446...87G,2006MNRAS.367..928V,2006ApJ...653.1115O,2008ApJ...684..811S,2009A&A...501...89T,2012MNRAS.424.1774H,2014MNRAS.437.3063K,2016ApJ...823...57S,2018A&A...612L...4B,2017ApJ...851...87O,2019ApJ...871...71S,2020ApJ...899..127S}. 
However, due to the intrinsic compactness that cause an unresolved X-ray morphology in most observations, the mechanism of X-ray emission in young AGN is still subject to controversy.
There are two main hypotheses about their X-ray emission. 
Observations found young radio AGNs tend to have X-ray spectral index, X-ray to OIII flux ratio, or X-ray to optical flux ratio similar to Seyfert galaxies, 
suggesting their X-ray emission be thermal inverse Comptonization process from an accretion disc$-$corona system \citep[e.g.,][]{2008ApJ...684..811S,2009A&A...501...89T}. Meanwhile, spectral energy distribution (SED) modeling suggests that the X-ray emission could probably be produced from non-thermal inverse Compton scattering processes in radio lobes for galaxies \citep{2008ApJ...680..911S,2010ApJ...715.1071O} or jets for quasars \citep{2012ApJ...749..107M,2014ApJ...780..165M}.  Moreover, recently works of statistic analysis by employing the fundamental plane of black hole activity, likely support the jet origin for a small sample with high-accretion rate including both young radio galaxies and quasars \citep{2016ApJ...818..185F,2020MNRAS.497..482L}.
While both mechanisms may contribute in individual sources, searching for direct observational evidence is essential to distinguish the two scenarios.

A prominent and key feature of the thermal coronal emission is the high energy cutoff ($E_{\rm cut}$) at $\sim$ 150 keV \citep{Molina}, which is a direct indicator of the coronal temperature $T_{\rm e}$. The  \emph{Nuclear Spectroscopic Telescope Array} \citep[NuSTAR,][]{2013ApJ...770..103H}, with its broad spectral coverage, for the first time provides good constraints on the $E_{\rm cut}$ in a number of sources \citep{2015A&A...577A..38U,Molina, Rani_2019, Panagiotou_2020, Kang_2020, Kang_2022}. 
While the high energy cutoff has been widely detected in the X-ray spectra of radio-quiet AGN, it is until recently that NuSTAR observations have detected the $E_{\rm cut}$ in a number of radio galaxies, mostly FR II galaxies \citep[e.g.][]{2014ApJ...794...62B,Kang_2020, Kang_2022}. This indicates the hard X-ray emission in regular radio galaxies (non-blazars) are dominated by coronal emission. 

The detection/measurement of $E_{\rm cut}$ in young radio AGN, is scientifically essential to explore whether their hard X-ray emission are dominated by coronal emission, likewise in regular radio galaxies, namely, to examine whether the physical properties of the central disc-corona-jet system change from the infant stage to the mature stage.

To date, we know little about the $E_{\rm cut}$ in young radio AGN because their X-ray emission is generally faint, making $E_{\rm cut}$ detection rather challenging \citep[e.g.,][]{Kang_2020,Kang_2022}. By cross-matching the parent sample of young radio AGN (including GPS, CSS, HFP and CSO) 
of \cite{2020MNRAS.491...92L} 
which is compiled from radio-selected samples in the literature, 
with NuSTAR data archive, we find 10 sources with NuSTAR observations. However all (but Mrk 348, the target of this work) of them have 3 -- 78 keV net counts (FPMA + FPMB) $<$ 6000, inadequate to constrain the $E_{\rm cut}$ in a typical AGN spectrum \citep{Kang_2020, Kang_2022}. 

Mrk 348 (NGC 262) is associated with a z = 0.015 Seyfert 2 galaxy with a hidden broad-line region showing the H$\alpha$ full width at half-maximum (FWHM) of 7400 $\rm km~s^{-1}$ from the polarization observations \citep{1990ApJ...355..456M}.
It hosts a supermassive black hole with mass of 6.3 $\times$ $10^{6}~\rm M_{\odot}$ derived using the stellar velocity dispersion, and is accreting in a radiatively efficient manner at high Eddington ratio \citep[$R_{\rm edd}$ = 0.56,][]{2022ApJS..261....2K}. 
Mrk 348 is a GPS radio source having radio spectrum peaked in the GHz range \citep[$\sim 4-5 \rm GHz$, ][]{2012ApJ...760...77A}. 
Multi-epoch high-resolution very long baseline interferometry (VLBI) observations 
\citep[e.g.][]{2003ApJ...590..149P} revealed that Mrk 348 contains a faded jet with double structures (extent of 180 mas/50 pc), and inner ($\sim$ 1pc at 15 GHz) evolved jet with two-sided morphology at similar flux densities \citep[i.e., CSO-like morphology which is also be considered in][]{2012ApJ...760...77A}. 
They inferred the separation speed of the inner jet between 0.22 and 0.34 mas $\rm yr^{-1}$. Together with the compact extent of the radio structure, this translates to an approximate kinematic age of 8-12 yr. This further indicates its radio axis is close to the plane of the sky and thus free from Doppler beaming effect. 

Mrk 348 is rather bright in X-ray with a NuSTAR count rate of 2.4 count/s, while the other nine sources are much fainter with count rates $<$ 0.4 count/s. 
The NuSTAR spectra of Mrk 348 with 21 ks exposure (OdsID 60160026002) have been presented in several works, however, with $E_{\rm cut}$ contrarily  constrained: \cite{Rani_2019} reports an $E_{\rm cut} = 80^{+40}_{-20}$ keV and \cite{Balokovi_2020} reports an $E_{\rm cut} = 170^{+40}_{-20}$ keV, while \cite{Panagiotou_2020} reports an $E_{\rm cut} >  270$ keV. The discrepancies among these works are likely  due to differences in the fitting procedures adopted. By simply adopting the model 2 in our work (see Section 3 for details), we could derive an $E_{\rm cut} > 150$ keV (90\% confidence range), generally consistent with \cite{Balokovi_2020} and \cite{Panagiotou_2020}. 

We obtained a new and much longer NuSTAR observation on Mrk 348 in 2022 February. The main goal of this study is to model its broad-band X-ray spectrum and measuring $E_{\rm cut}$, thus investigate the origin of its X-ray emission, as well as to directly compare its X-ray properties with large-scale radio galaxies to explore the evolution of X-ray emission. 


The outline of our work is as follows: \S2 presents data reduction; \S3 presents the X-ray spectral analysis; \S4 provides the results and discussion, where we do not involve the earlier studies with quasars' counterparts of young radio AGN, unless otherwise specified, which may have significant beaming effect and the inclusion of which would bias larger contribution from the jets to the X-ray emission.
Throughout the paper, the cosmological parameters $H_0 = 70\, \mathrm{km\,s^{-1}\, Mpc^{-1}}$, $\Omega_\mathrm{m} = 0.3$, and $\Omega_{\lambda} = 0.7$ are adopted.


\section{New NuSTAR OBSERVATION AND DATA REDUCTION}

\par Mrk 348 was observed by NuSTAR on 2022-02-06 with an exposure time of 105 ks (ObsID=60701017002). The data reduction is performed by employing NuSTAR Data Analysis Software (NuSTARDAS) in HEASoft package (version 6.28), with CALDB version 20201101 for calibration. We use  \emph{nupipeline} to generate the calibrated and cleaned event files and \emph{nuproduct} to extract the source spectra within a circle with a radius of 60$^{''}$ centered on our target coordinate (00h48m47.1414s, +31d57m25.085s). We use NUSKYBGD \citep{Wik_2014} for background estimation, which simulates a background spectrum within the source region by modelling the background all over the field of view (FOV), taking care of the spatial variation of the background \citep[see Figure 1 in][for example]{Kang_2020}. Furthermore, to utilize the $\chi^{2}$ statistics for spectral fitting, we re-bin the spectra with \emph{grppha} to guarantee a minimum of 100 counts per bin. We shall investigate how the background estimation and spectral binning affect the measurement of spectral parameters in \S \ref{S:first_model}.



\begin{figure*}
	\includegraphics[width=0.95\textwidth]{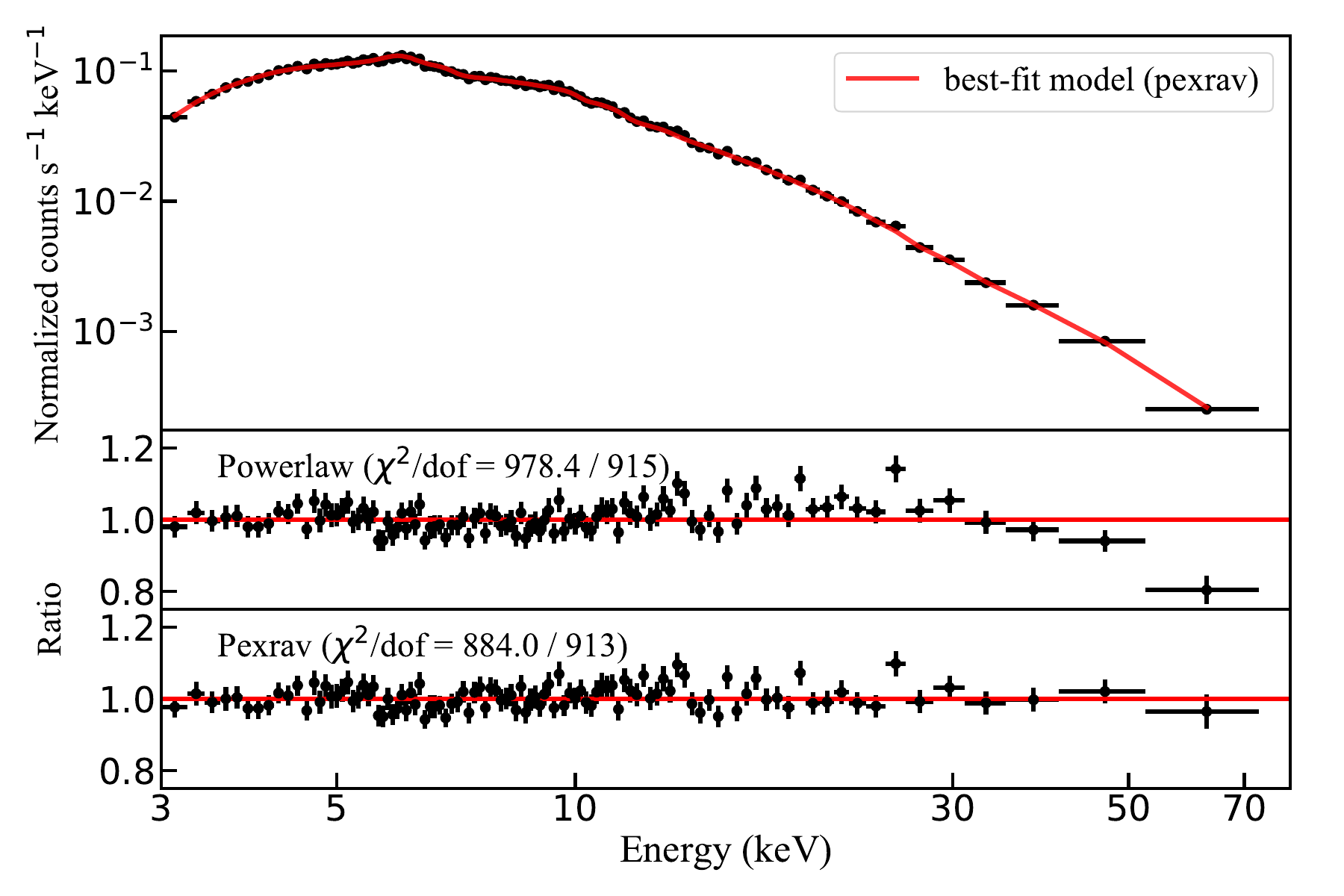}
	\caption{Upper panel: the NuSTAR spectrum along with the best-fitting model of \emph{pexrav} (an exponentially cutoff power-law with a neutral reflection component) plus $zgauss$; note the error bars of the data are too small to be noticed in the figure. Middle panel: the data to model ratio when fitting the NuSTAR spectrum with a single \emph{powerlaw} plus \emph{zgauss}. Lower panel: the data to model ratio of \emph{pexrav} plus $zgauss$. For better illustration, only the spectrum of FPMA is plotted after further re-binning, while the best-fitting models are derived by jointly fitting the spectra of FPMA and FPMB.}
 \label{fig:spectra_fitting}
\end{figure*} 

\section{SPECTRAL ANALYSES}
We perform NuSTAR spectral analyses and model fitting in the 3-79 keV band with XSPEC package \citep{1996ASPC..101...17A}, using $\chi^{2}$  minimization technique. Here we list the spectral models (including both phenomenological and physical ones) which are used later:\\
\\
(1) $constant*zphabs*(powerlaw+zgauss)$\\
(2) $constant*zphabs*(pexrav+zgauss)$\\
(3) $constant*zphabs*relxilllpCp$\\
(4) $constant*zphabs*(nthComp+pexrav+zgauss)$\\
(5) $constant*zphabs*(compPS+zgauss)$\\
(6) $constant*zphabs*(compTT+pexrav+zgauss)$\\

In all the models we adopted, the multiplicative \emph{constant} model is included to account for the difference in flux calibration between instruments, and \emph{zphabs} the neutral intrinsic absorption. Galactic absorption is ignored due to its little impact on NuSTAR spectra. The \emph{constant} is fixed at 1.0 for NuSTAR/FPMA, and allowed to vary for NuSTAR/FPMB. The neutral intrinsic absorption $N_{H}$ is kept free during the fit for each model, unless otherwise stated.
The \emph{zgauss} models the Fe K$\alpha$ line. Unless otherwise specified, its centroid energy, width and normalization are all allowed to vary if it's involved.
The errors of the spectral fitting parameters and of any other derived quantities are at the 90$\%$ confidence level ($\rm \Delta \chi^{2} = 2.71$) for one interesting parameter.

\subsection{Phenomenological models} \label{S:first_model}
We initially fit the 2022 NuSTAR spectra with a simple power-law  plus \emph{zgauss} (model 1), to test whether the presence of a curvature at high energy. 
We obtain a moderate fit ($\chi^{2}$/dof = 978.4/915) with apparent descent above 50 keV (see Fig. \ref{fig:spectra_fitting}). To reproduce this curvature, in model 2 we replace the \emph{powerlaw} of model 1 with \emph{pexrav} \citep{1995MNRAS.273..837M}, which is extensively used by the community \citep[e.g.,][] {Molina,Panagiotou_2020}. \emph{pexrav} describes the exponentially cutoff power-law plus the Compton reflection component off a neutral slab geometry of infinite column density. In such a fitting process, we fix the inclination cos$i$ = 0.45 that is appropriate for a type 2 AGN, assume solar abundances, and leave free the reflection fraction $R$ = $\Omega$/2$\pi$, $\Gamma$, and $E_{\rm cut}$ and $N_{\rm pexrav}$. We obtain $E_{\rm cut}$ = $218^{+124}_{-62}$ keV with a significantly improved fit ($\chi^{2}$/dof = 884.0/913). Moreover, as shown by the $\Gamma$--\ec contour in Figure \ref{fig:contour}, \ec could be well constrained at 99\% confidence level even after considering the degeneracy between $\Gamma$ and $E_{\rm cut}$.
Meanwhile, the best fitting results of model 2 yield the Fe k$\alpha$ line center at $6.09^{+0.07}_{-0.09}$ keV with $\sigma = 0.23^{+0.12}_{-0.10}$ keV and EW = $106^{+37}_{-26}$ keV. The significant broadening and redshifted centroid energy of the line indicate it may originate from the relativistic reflection from the inner disc where the effects of relativistic Doppler and gravitational redshift are at work. 

\par {To examine the robustness of the measured $E_{\rm cut}$, we inspect the influence of background estimation and spectral binning. Besides estimating the background with NUSKYBGD, we also extract the background spectra simply from a source-free circular region (with a radius of 60\arcsec) nearby the source. As shown in Table \ref{tab:pexrav}, compared with NUSKYBGD ($E_{\rm cut}$ = $218^{+124}_{-62}$ keV), utilizing the background spectra from a circular region yields slightly smaller but statistically consistent $E_{\rm cut}$ = $200^{+101}_{-56}$ keV. Considering \ec measurements are sensitive to the high energy data where the background is more dominant, using NUSKYBGD to obtain a more accurate estimation of the background is important and adopted hereafter.   

\par {Meanwhile, we try different spectral binning criteria, by requiring a minimum of 20, 50 and 100 counts per bin, respectively. As shown in Table \ref{tab:pexrav}, the measured $E_{\rm cut}$ (keV)  are  $180^{+77}_{-47}$, $203^{+105}_{-55}$, and $218^{+124}_{-62}$, respectively; the smaller the bin, the smaller the $E_{\rm cut}$. We then perform simulations to investigate whether such discrepancy is systematic. For saving the computing time, the simpler \emph{cutoffpl} model, a power-law continuum with a high-energy cutoff (without a reflection component), is adopted as input, with $\Gamma$ set at 1.6 and \ec at 220 keV (approximately the best-fit values of \emph{pexrav}). Using the input model and NuSTAR response files, we create 1000 pairs of mock spectra (of FPMA and FPMB) with the \emph{fakeit} task within XSPEC, which have similar count rates ($\sim$ 1.0 counts/s for FPMA) and background level to those of the observed spectra.  We then re-bin these mock spectra to achieve a minimum of 20, 50 and 100 counts per bin respectively, fit the re-binned spectra with a \emph{cutoffpl}, and measure the $E_{\rm cut}$ \footnote{The signal-to-noise ratio is high enough to constrain the \ec for all the mock spectra at 90\% confidence level, so the percentiles in Figure \ref{fig:fake} could be directly derived without considering lower limits.}. As shown in Figure \ref{fig:fake}, the spectral binning does significantly influence the measurement of \ec; particularly, requiring a minimum of 20 or 50 counts per bin would systematically underestimate \ec, while a minimum of 100 counts per bin (adopted in this work) could well reproduce the input $E_{\rm cut}$. 

\par {To summarize, the 
spectral binning does affect the measurement of $E_{\rm cut}$. In this work, we estimate the background with NUSKYBGD and bin the spectra to achieve a minimum of 100 counts per bin, which lead to more accurate and conservative measurement of \ec than other widely accepted approaches (see Table \ref{tab:pexrav}). 

\begin{figure}
	\includegraphics[width=0.45\textwidth]{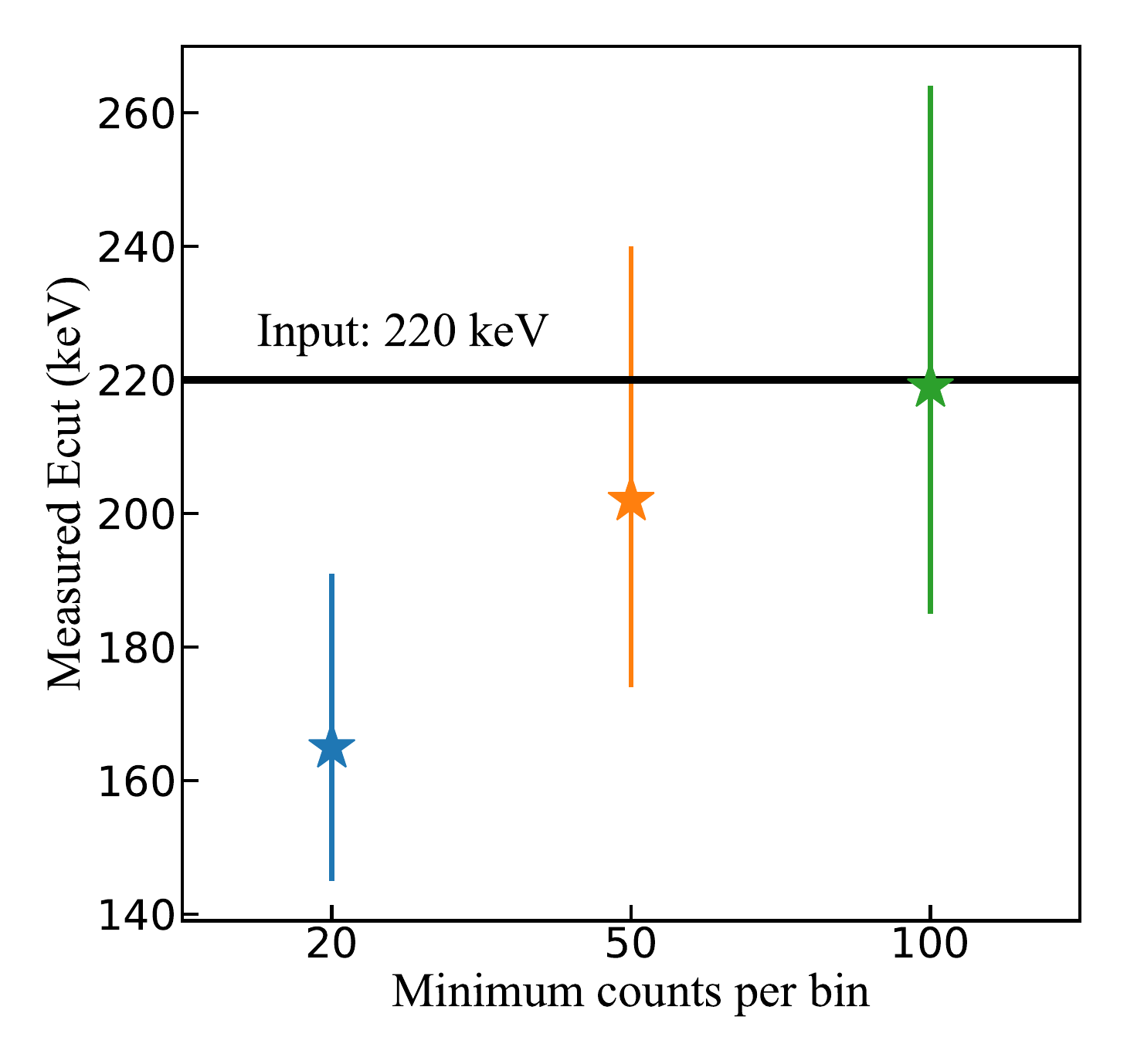}
	\caption{The median value of \ec for the mock spectra with different binning criteria; error bars represent the 1$\sigma$ scatter (16th to 84th percentiles). } 
 \label{fig:fake}
\end{figure} 


\begin{table*}
\large
	\centering
	\caption{Best-fit spectral parameters derived from model 2 ($pexrav$), for different methods of background estimation and spectral binning. }\label{tab:pexrav}
\begin{tabular}{ccccc} 
\hline
Background method & NUSKYBGD &  NUSKYBGD & Nearby region & NUSKYBGD\\
\hline
Minimum counts bin$^{-1}$ & 25 & 50 & 100 & 100\\
\hline
\hline
$N_{\rm H}$ ($ \rm 10^{22} cm^{-2} $) & $8.2^{+0.6}_{-0.7}$ & $8.2^{+0.6}_{-0.7}$ & $8.2^{+0.6}_{-0.7}$ & $8.2^{+0.6}_{-0.7}$\\
$\Gamma$ & $1.58^{+0.06}_{-0.07}$ & $1.59^{+0.06}_{-0.07}$ & $1.59^{+0.06}_{-0.07}$ & $1.59^{+0.06}_{-0.07}$\\
$E_{\rm cut}$ (keV) & $180^{+77}_{-47}$ & $203^{+105}_{-55}$ & $200^{+101}_{-56}$ & $218^{+124}_{-62}$\\
$R$ & $0.36^{+0.11}_{-0.11}$ & $0.38^{+0.11}_{-0.11}$ & $0.40^{+0.11}_{-0.11}$ & $0.39^{+0.11}_{-0.10}$\\
$E_{\rm c}$ (keV) & $6.09^{+0.07}_{-0.10}$ & $6.09^{+0.07}_{-0.09}$ & $6.08^{+0.07}_{-0.10}$ & $6.09^{+0.07}_{-0.09}$\\
$\sigma$ (keV) & $0.23^{+0.13}_{-0.10}$ & $0.23^{+0.12}_{-0.10}$ & $0.23^{+0.13}_{-0.10}$ & $0.23^{+0.12}_{-0.10}$\\
EW (eV) & $106^{+39}_{-25}$ & $105^{+37}_{-26}$ & $105^{+38}_{-25}$ & $106^{+37}_{-26}$\\
$\chi^2$/dof & 1467.9/1517 & 1145.8/1193 & 880.3/913 & 884.0/913\\
\hline
\end{tabular}
\end{table*}

\begin{table}
\large
	\centering
	\caption{Best-fit spectral parameters derived from $relxillpCp$, $nthComp$, $compPS$ and $compTT$, for the NuSTAR observation obtained in 2022.}\label{tab:results}
\begin{tabular}{cc} 
\hline
Parameter & value\\
\hline
\hline
\multicolumn{2}{c}{Model 3 ($relxilllpCp$)}\\
\hline
$N_{\rm H}$ ($ \rm 10^{22} cm^{-2} $) & $8.8^{+0.4}_{-0.3}$ \\
Inclination (degree) & $3^{+-3}_{-3}$ \\
$h\,(R_{\rm g})$ & $16^{+5}_{-3}$ \\
$Afe$ & $3.6^{+0.8}_{-1.0}$ \\
$\Gamma$ & $1.68^{+0.03}_{-0.02}$ \\
$kT_{\rm e}$ (keV) & $36^{+14}_{-7}$ \\
$R$ & $0.31^{+0.08}_{-0.03}$ \\
$\chi^2$/dof & 893.1/913 \\
\hline
\multicolumn{2}{c}{Model 4 ($nthComp$)}\\
\hline
$N_{\rm H}$ ($ \rm 10^{22} cm^{-2} $) & $8.9^{+0.3}_{-0.3}$ \\
$\Gamma$ & $1.70^{+0.01}_{-0.01}$ \\
$kT_{\rm e}$ (keV) & $40^{+44}_{-10}$ \\
$E_{\rm c}$ (keV) & $6.12^{+0.05}_{-0.06}$ \\
$\sigma$ (keV) & $0.17^{+0.08}_{-0.08}$ \\
$\chi^2$/dof & 887.4/914 \\
\hline
\multicolumn{2}{c}{Model 5 ($compPS$)}\\
\hline
$N_{\rm H}$ ($ \rm 10^{22} cm^{-2} $) & $8.3^{+0.7}_{-0.6}$ \\
$kT_{\rm e}$ (keV) & $35^{+7}_{-5}$ \\
Compton $y$ & $0.61^{+0.03}_{-0.02}$ \\
$R$ & $0.28^{+0.13}_{-0.10}$ \\
$E_{\rm c}$ (keV) & $6.08^{+0.08}_{-0.09}$ \\
$\sigma$ (keV) & $0.20^{+0.15}_{-0.08}$ \\
$\chi^2$/dof & 886.7/913 \\
\hline
\multicolumn{2}{c}{Model 6 ($compTT$)}\\
\hline
$N_{\rm H}$ ($ \rm 10^{22} cm^{-2} $) & $8.2^{+0.4}_{-0.4}$ \\
$kT_{\rm e}$ (keV) & $94^{+199}_{-67}$ \\
$\tau$ & $0.56^{+1.49}_{-0.41}$ \\
$E_{\rm c}$ (keV) & $6.09^{+0.06}_{-0.08}$ \\
$\sigma$ (keV) & $0.22^{+0.10}_{-0.09}$ \\
$\chi^2$/dof & 884.0/914 \\
\hline
\end{tabular}
\end{table}

\begin{figure}
	\includegraphics[width=3.4in]{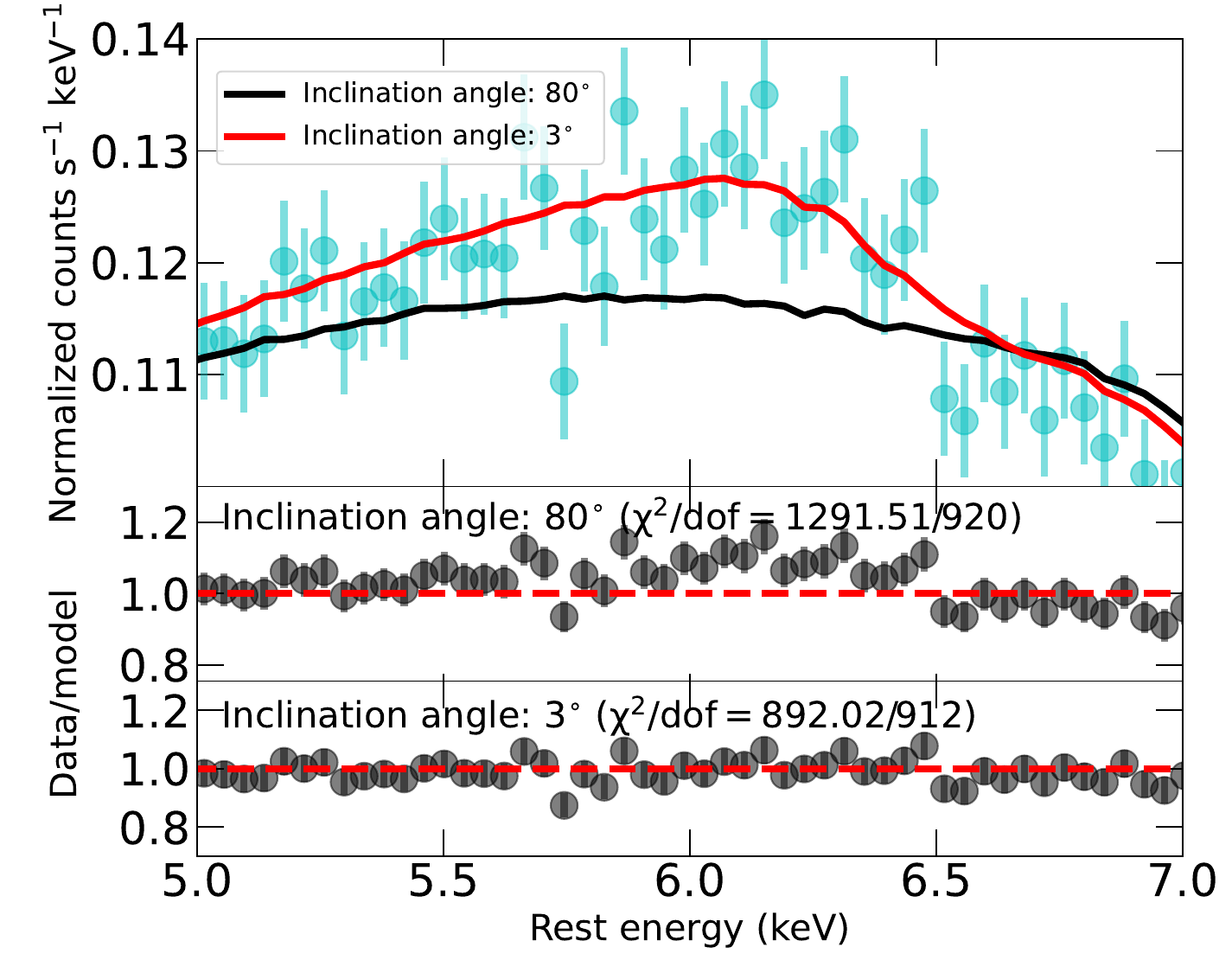}

	\caption{The fitting of Fe K$\alpha$ line profile with $\theta_{\rm incl}$ =  $3^{\circ}$ (best-fit) and $80^{\circ}$ (fixed)
 under the best-fit results from model (3) 
 using \emph{relxilllpCp}. 
 Evidently, the residual is present when $\theta_{\rm incl}$ is at $80^{\circ}$.
    } \label{fig:inclination}
\end{figure}

\subsection{Physical models}
The high S/N of our NuSTAR spectra over 3-79 keV enable us to consider more physically motivated models to describe its continuum and line emission and directly measure the corona temperature ($kT_{\rm e}$). In model 3 we replace the \emph{pexrav} and \emph{zgauss} with the lamp-post model \emph{relxilllpCp} \citep{2013ApJ...768..146G,2014ApJ...782...76G}, which describes a thermal Comptonization continuum $nthcomp$ associated with $kT_{\rm e}$, a relativistically blurred reflection off a disc, and meanwhile self-consistently incorporates the Fe K$\alpha$ line. In this model, we consider a maximally spinning black hole with the spin parameter fixed to 0.998 (the fit is found to
be quite insensitive to the spin). The disk ionization log $\xi$ is fixed at zero, and  
the inner ($R_{\rm in}$) and outer radius ($R_{\rm out}$) of the accretion disk at 1 ISCO and 400 $R_{\rm g}$, respectively. 
This model yields a fit with $\chi^{2}$/dof = 893.1/913 and $kT_{\rm e}$ = $36^{+14}_{-7}$ keV. 
However, we obtain $\theta_{\rm incl}$ $<$ $10^{\circ}$ which requires a face-on orientation to the disk, obviously contradicting the edge-on geometry inferred from optical and radio observations in Mrk 348, as aforementioned \citep[e.g.,][]{1990ApJ...355..456M,2003ApJ...590..149P,2012ApJ...760...77A}.
The fitting to the Fe K line profile would be rather poor if we adopt an edge-on geometry (e.g., by fixing the $\theta_{\rm incl}$ at $80^{\circ}$, 
as plotted in Fig. \ref{fig:inclination}).
Allowing the parameters such as spin $a$, disk ionization log $\xi$, and inner radius $R_{\rm in}$ to vary can not ease the tension. 
The discrepancy is because the bulk of the Fe K line emission is significantly redshifted, which can only be attributed to strong gravitational redshift from the inner disk but only in a face-on geometry, as an edge-on geometry would yield significant blueshifted component because of the Doppler shift/beaming effect. 
This hints that the Fe K line of Mrk 348 can not be explained by the simple lamp-post model.


\begin{figure*}
\centering
\subfloat{\includegraphics[width=0.34\textwidth]{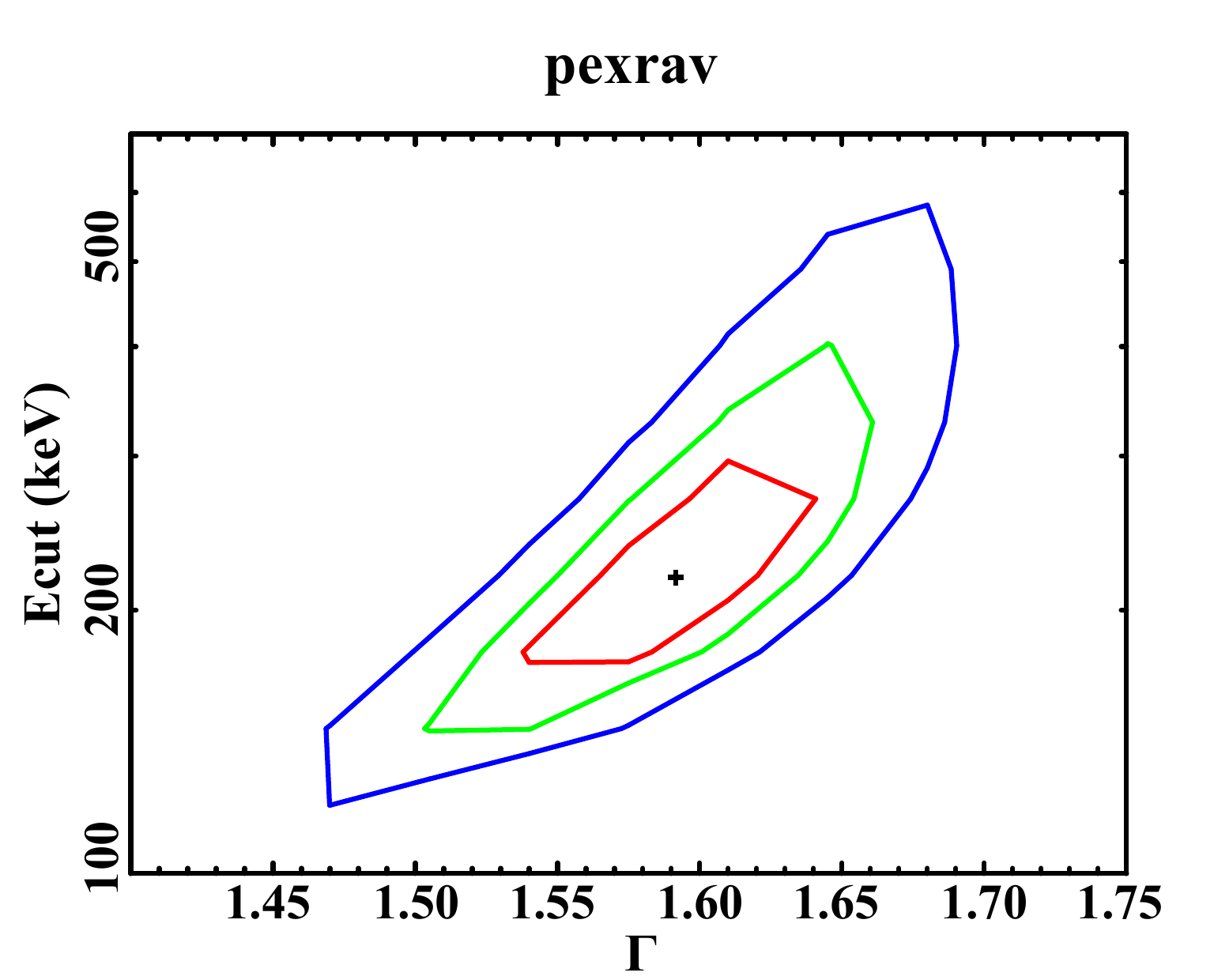} }\hspace{-0.5cm}
\subfloat{\includegraphics[width=0.34\textwidth]{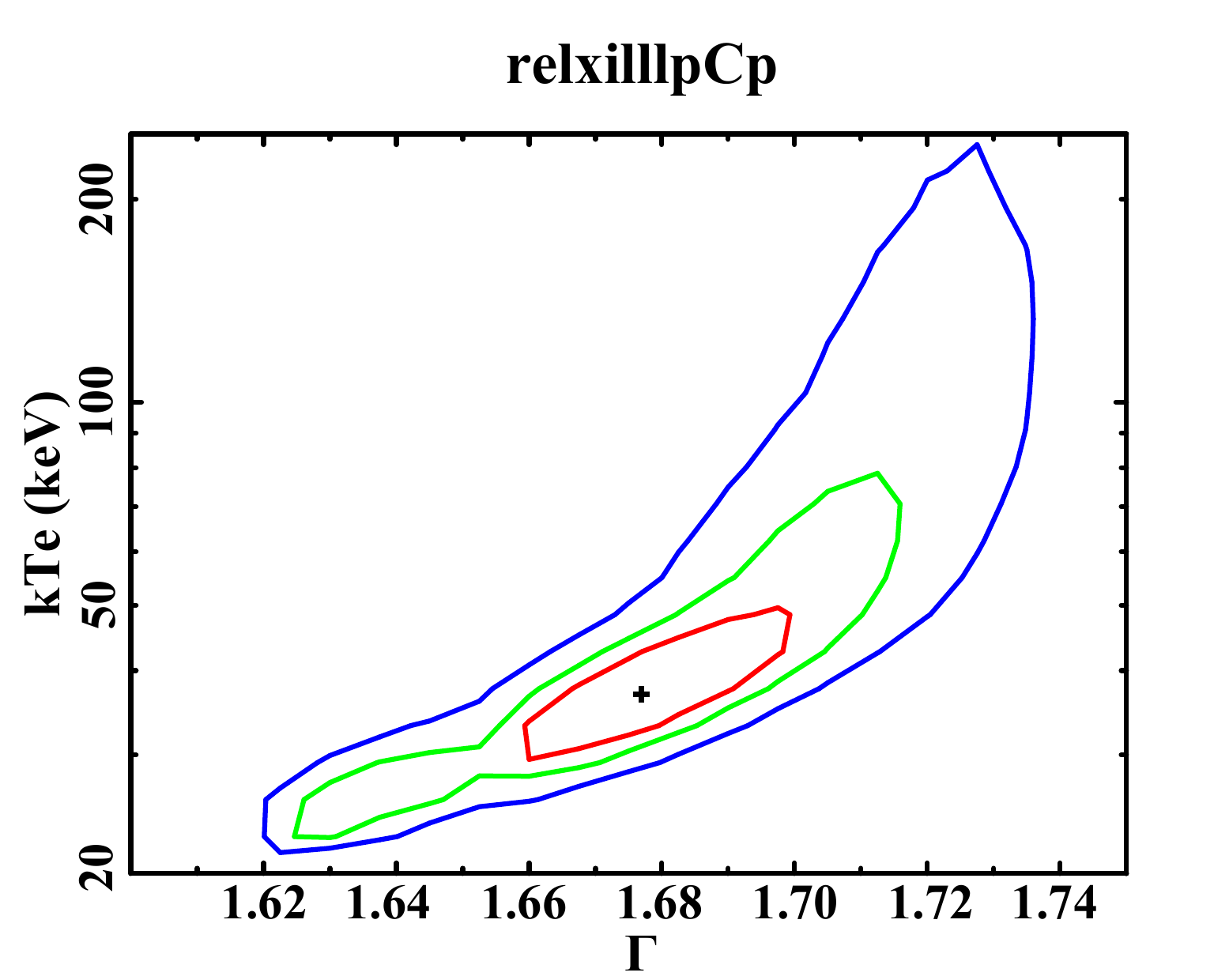} }\hspace{-0.5cm}
\subfloat{\includegraphics[width=0.34\textwidth]{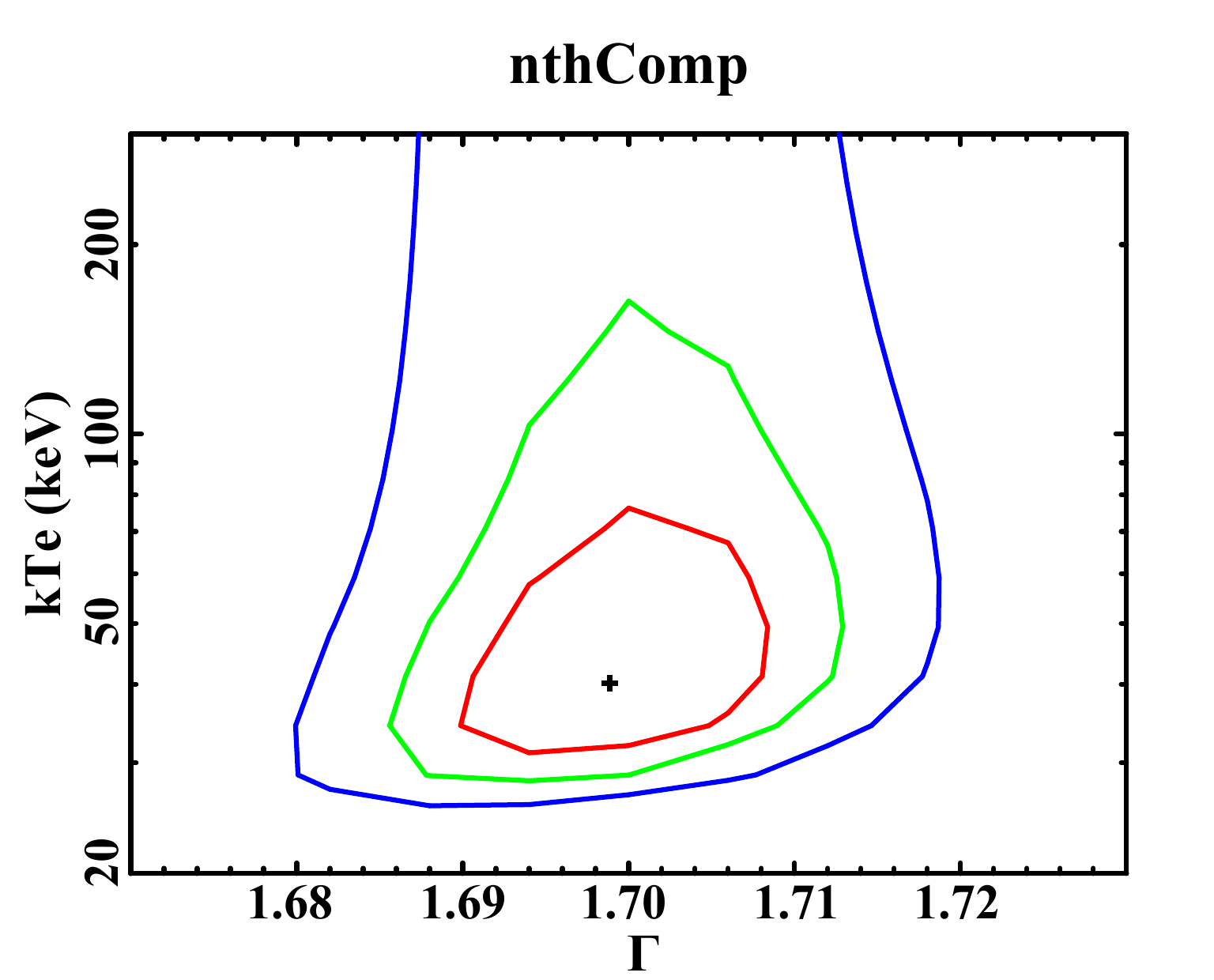} }\vspace{-0.5cm}\\
\subfloat{\includegraphics[width=0.34\textwidth]{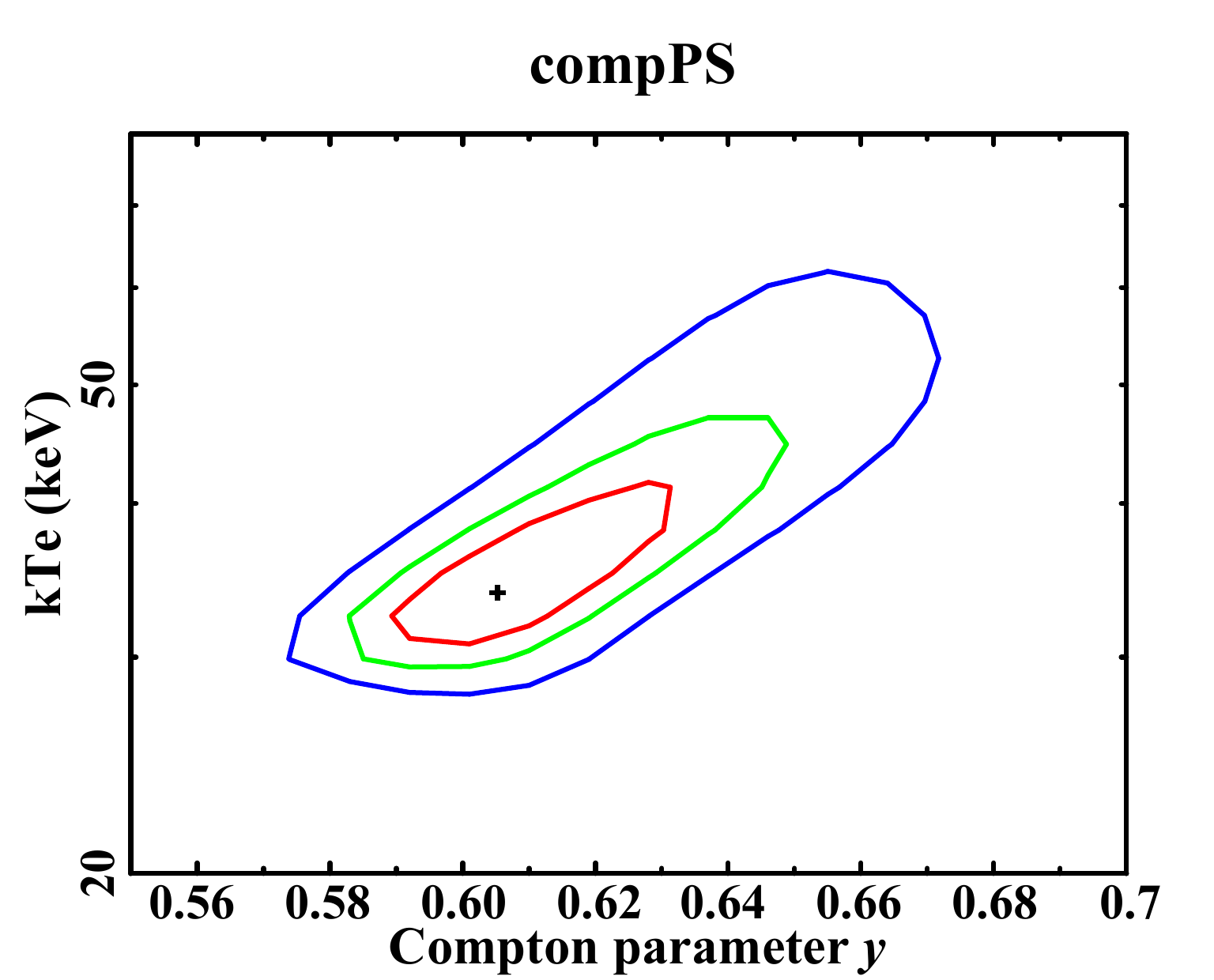} }\hspace{-0.5cm}
\subfloat{\includegraphics[width=0.34\textwidth]{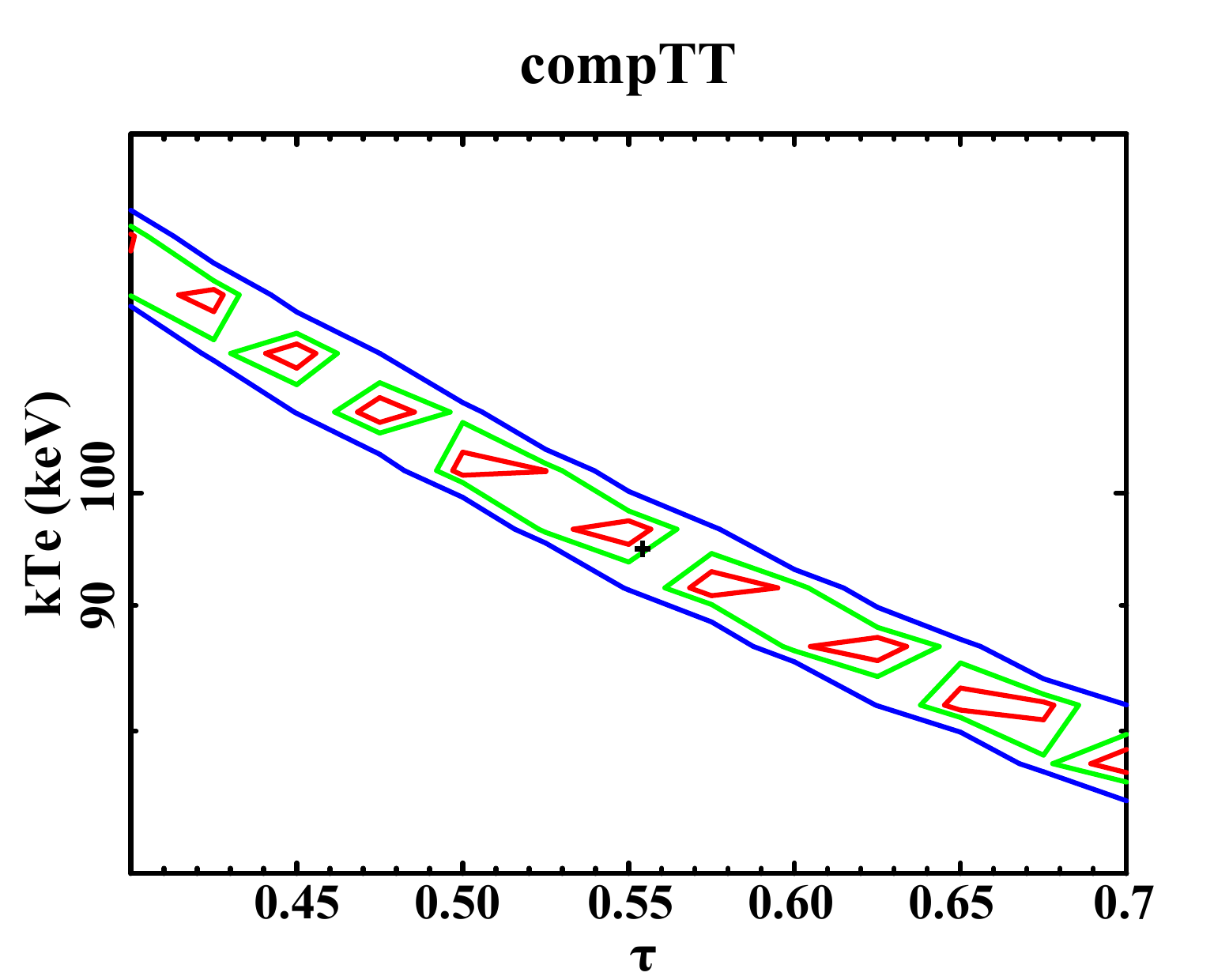} }
\caption{\label{fig:contour} The two-dimension contours of $E_{\rm cut}$/$kT_{\rm e}$ versus the other key parameter of the continuum, for model 2--6. The red, green and blue curves represent the 68\%, 90\% and 99\% confidence level ($\Delta \chi^2$ = 2.30, 4.61 and 9.21), while the black cross marks the best-fit value.   
}
\end{figure*}

\par {We then adopt another model (model 4), \emph{nthComp} \citep{Zdziarski_1996, Zycki_1999}, which describes a thermally Comptonized continuum. Since neither a Fe line nor a reflection component is included in \emph{nthComp}, we keep \emph{zgauss} to model the Fe K$\alpha$ line and fix the reflection component at that derived by model 2 (\emph{pexrav}), by setting the $R$ in \emph{pexrav} to its additive inverse and freeze all the parameters of \emph{pexrav}. As for \emph{nthComp}, the seed photons are set to follow a blackbody distribution with temperature $kT_{\rm bb} = $ 0.1 keV, while $\Gamma$ and $kT_{\rm e}$ are allowed free to vary during the fitting. This model, referred as model 4, derives a $kT_{\rm e}$ = $40^{+44}_{-10}$ keV. As shown in Figure \ref{fig:contour}, $kT_{\rm e}$ is well constrained at 90\% confidence level, while unconstrained at 99\% confidence level in the $\Gamma$-$kT_{\rm e}$ contour. We note the practice of fitting with a fixed reflection component may affect the measurement of $E_{\rm cut}$/$kT_{\rm e}$, as the reflection component is sensitive to the spectral shape of the hardest coronal radiation \citep[e.g.,][]{ Kang_2022}. 

We further employ the Comptonization model of \emph{compPS}  \citep{1996ApJ...470..249P} to replace the \emph{pexrav} of model 2, as expressed in model 5. 
We fix the temperature of soft photons at $kT_{\rm bb}$ = 10 eV, cos$i$ = 0.45, and allow the $kT_{\rm e}$,  the Compton parameter $y$ (= $4\tau kT_{\rm e}/m_{\rm e}c^{2}$, where $\tau$ is the optical depth), and $R$ (the amount of reflection) 
free to vary. Other not mentioned parameters are fixed at the default values of the model. We consider a slab geometry for the hot corona, which is favored by recent X-ray polarization observation of AGN \citep[g.g.][]{Gianolli2023}.
Model 5 yields $\chi^{2}$/dof = 886.7/913, $kT_{\rm e}$ = $35^{+7}_{-5}$ keV, and $y$ = $0.61^{+0.03}_{-0.02}$. As shown in Figure \ref{fig:contour}, $kT_{\rm e}$ and $y$ could be well constrained at 99\% confidence level after considering the degeneracy between them. 
Moreover, the resulted values of $R$, $N_{\rm H}$, and the Fe K energy centroid energy/width agree well with model 2 within statistical errors.

\par {Finally, we adopt another widely used Comptonization model, \emph{compTT} \citep{Titarchuk_1994}. Similar to model 4 (\emph{nthComp}), we keep \emph{pexrav} and \emph{zgauss} to model the reflection component and Fe K$\alpha$ line, respectively, shown as model 6. For \emph{compTT}, we suppose an input seed photon (Wien) temperature $T_{0}$ of 0.1 keV and a disk geometry of the corona. This model obtains a marginal detection of $kT_{\rm e}$ ($ = 94^{+199}_{-67}$ keV). However, as shown in Figure \ref{fig:contour}, the degeneracy between $kT_{\rm e}$ and optical depth $\tau$ is extremely strong, making it difficult to determine a global best-fit. Beside, adopting a different seed photon temperature or a sphere geometry of the corona do not improve the constraint of $kT_{\rm e}$. 
}

\par {In conclusion, we adopt four different physical models to fit the NuSTAR spectra, three of which put robust and consistent constraint on $kT_{\rm e}$ ($ \sim$ 35--40 keV). The fitting results of these models are shown in Table \ref{tab:results}, with the two dimensional contours of their key parameters in Figure \ref{fig:contour}.

\section{Discussion}

In this study, for the first time, we report the detection of the X-ray high energy cutoff ($E_{\rm cut}$ = $218^{+124}_{-62}$ keV, \emph{pexrav}) and the measurement of corona temperature (35 -- 40 keV) in a young radio AGN Mrk 348.
The clear detection of $E_{\rm cut}$ strongly indicates that the thermal corona dominates the hard X-ray emission in Mrk 348. 





\begin{figure}
	\includegraphics[width=3.4in]{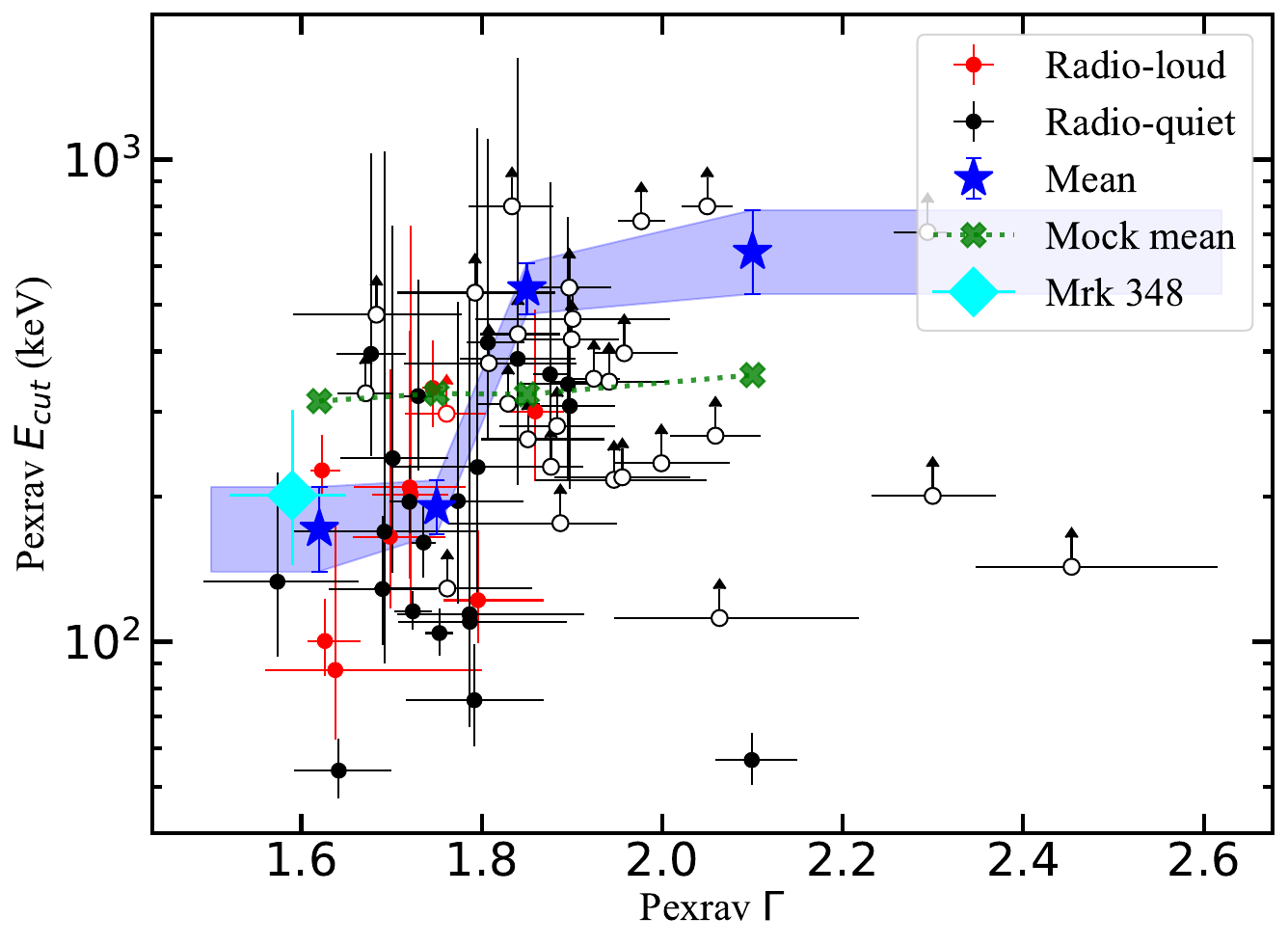}
	\caption{Mrk 348 lies on the $E_{\rm cut}$ - $\Gamma$ track of \citet{Kang_2022}.} \label{Fig:Kang_comparisons}
\end{figure}


Previous studies have shown the X-ray properties throughout the evolution of radio AGN, may primarily depend on the AGN accretion. 
\cite{2014MNRAS.437.3063K} find that HEG and LEGs galaxies, usually associated with high
accretion rate and low accretion rate, respectively, occupy a distinct branches in the radio/X-ray luminosity plane, notwithstanding their evolutionary stage, (i.e., hold for both young radio AGN and FRII/FRI radio galaxies), suggesting a different origin of the X-ray emission for different accretion level, namely, disc-corona for HEGs, while jet base for LEGs. 
\cite{Kang_2020} presented the first systematic study of NuSTAR hard X-ray spectra for radio-loud AGNs. Their sample includes 20 FR II and 8 FR I. Among them 13 sources have the $E_{\rm cut}$ detection and the non-detections could be attributed to too low NuSTAR counts, implying the X-ray emission in non-blazar radio AGN is dominated by corona but not the jet (though significant jet contamination to their FR I or core-dominated sources could not be ruled out). 
The NuSTAR detection of $E_{\rm cut}$ in FR II \citep{Kang_2020} seems support the scenario that the X-ray emission of the radio sources with high-accretion level is corona dominated. 
However, the $E_{\rm cut}$ has also been detected in a few FR I radio galaxies, implying the same corona origin of X-ray emission in at least some FR I \citep{Kang_2020}. 
We note Mrk 348 does have a high Eddington ratio of 0.56 \citep{2022ApJS..261....2K}. Thus, the $E_{\rm cut}$ detection in Mrk 348 together with those in FR II, fit well the scenario that accretion-corona system dominates X-ray emission in high-accretion objects.  
On the other hand, it has been suggested that
the maturity as FR II or FR I radio galaxies depends on the radio power of young radio AGN \citep[more powerful young radio AGN firstly evolved into FR II and then into FR I, while less powerful ones would directly evolve
into FR I, ][]{2012ApJ...760...77A}.
Mrk 348 a low-power young radio AGN with log $L_{\rm 1.4GHz}$ = 24.53 ($\rm W~Hz^{-1}$) which would directly evolve into FR I based on the radio evolution pattern \citep{2012ApJ...760...77A}. { The results of this work thus suggests the X-ray emission in the infant stage with high-accretion of a FR I galaxy could also be dominated by the corona.}

\cite{Kang_2022} further found radio loud AGN tend to have lower $E_{\rm cut}$ compared with radio quiet ones, which could be attributed to the positive correlation between $E_{\rm cut}$ and $\Gamma$ their discovered.
In Fig. \ref{Fig:Kang_comparisons} we mark Mrk 348 in the plane of $E_{\rm cut}$ versus $\Gamma$, 
comparing with the NuSTAR bright radio quiet and loud AGN of \cite{Kang_2022}.
Clearly, Mrk 348 lies on the same $E_{\rm cut}$ -- $\Gamma$ track as both radio loud and radio quiet AGN. 
This indicates the corona of Mrk 348 is similar to other AGN. If this trend could be confirmed in large sample of young radio AGN, it suggests the X-ray corona properties and its dominance to X-ray emission of (non-blazar) radio AGN
may not significantly evolve from the infant stage (young radio AGN) to the mature stage.

We conclude that, from the NuSTAR perspective with its outstanding broad energy coverage, we provide the first direct observational evidence of corona dominated X-ray emission in a GPS/CSO galaxy. 
This suggests possible invariance of the X-ray emission mechanism throughout the evolution of radio AGN (i.e., persisting the corona-dominated thermal Comptonization emission), which is to be confirmed through extending this study of an individual source to a large sample of young radio AGN. Moreover, in future studies the inclusion of CSS radio sources (i.e., medium-scaled radio galaxies), and young radio AGN at low level of accretion, would be intriguing to shed light on the full characterization of X-ray emission across the radio AGN evolution. }

\section*{Acknowledgements}
We thank the anonymous referee for constructive suggestions which is great helpful in improving the manuscript.
We thank Minfeng Gu, Tao An for useful discussions. 
This work was supported by the National Science Foundation of China (Grants No. 12033006 $\&$ 12192221), and the Cyrus Chun Ying Tang Foundations. ML is supported the International Partnership Program of Chinese Academy of Sciences, Program No.114A11KYSB20210010.
MHZ is supported by the Natural Science Foundation of Jiangxi (No. 20232BAB211024), and the Doctoral Scientific Research Foundation of Shangrao Normal University (Grant No. K6000449).
XFL is supported by the National Science Foundation for Young Scientists of China (12203014)

This work made use of data from the NuSTAR mission, a project led by the
California Institute of Technology, managed by the Jet Propulsion Laboratory, and funded by the National Aeronautics and Space Administration. We thank the
NuSTAR Operations, Software, and Calibration teams for support with the execution and analysis of these observations. This research has made use of the
NuSTAR Data Analysis Software (NuSTARDAS) jointly developed by the ASI
Science Data Center (ASDC, Italy) and the California Institute of Technology
(USA). This research has made use of data and/or software provided
by the High Energy Astrophysics Science Archive Research Center (HEASARC), a
service of the Astrophysics Science Division at NASA/GSFC and of the
Smithsonian Astrophysical Observatory’s High Energy Astrophysics Division. This research has made use of data obtained from the Suzaku satellite, a collaborative mission between the space agencies of Japan (JAXA) and the USA
(NASA). This paper made use of data from XMM-Newton, an ESA science mission with instruments and contributions directly funded by the ESA
Member States and NASA. This research has made use of SAOImageDS9, developed by Smithsonian Astrophysical Observatory"
The National Radio Astronomy Observatory is a facility of the National Science Foundation operated under cooperative agreement by Associated Universities, Inc.

\section*{Data Availability}
This work is based on public X-ray data available from NuSTAR, XMM, Chandra, Suzaku.

\nocite{*}
\bibliographystyle{mnras}
\bibliography{ms} 




\bsp	
\label{lastpage}
\end{document}